# SPIDER: Specification-based Integration Defect Revealer

Vladislav Feofilaktov[1,2][0000-0002-2832-1661] and Vladimir Itsykson[1,2][0000-0003-0276-4517]

[1] Peter the Great St. Petersburg Polytechnic University, St. Petersburg, Russia
[2] JetBrains Research, Russia
vladislavf7@gmail.com, vlad@icc.spbstu.ru

**Abstract.** Modern software design practice implies widespread use in the development of ready-made components, usually designed as external libraries. The undoubted advantages of reusing third-party code can be offset by integration errors that appear in the developed software. The reason for the appearance of such errors is mainly due to misunderstanding or incomplete understanding by the programmer of the details of external libraries such as an internal structure and the subtleties of functioning. The documentation provided with the libraries is often very sparse and describes only the main intended scenarios for the interaction of the program and the library. In this paper, we propose the approach based on the use of formal library specifications, which allows detecting integration errors using static analysis methods. To do this, the external library is described using the LibSL specification language, the resulting description is translated into the internal data structures of the KEX analyzer. The execution of the incorrect scenarios of library usage, such as the incorrect sequence of method calls or the violation of the API function contract, is marked in the program model with special built-in functions of the KEX analyzer. Later, when analyzing the program, KEX becomes able to detect integration errors, since incorrect library usage scenarios are diagnosed as calling marked functions. The proposed approach is implemented as SPIDER (SPecification-based Integration Defect Revealer), which is an extension of the Kex analyzer and has proven its efficiency by detecting integration errors of different classes on several special-made projects, as well as on several projects taken from open repositories.

## 1  Introduction

Modern software is usually not designed from scratch. In most cases, when designing software, the developer actively uses external components, implemented, most often, as libraries. This approach has obvious advantages: you can reduce design and development time by reusing other people's solutions that have successfully proven themselves in other projects. However, in the case of using large libraries with the complex internal structure and behavior, integration errors may appear. They are caused by the developer's misunderstanding or incomplete understanding of the features of the library organization or misunderstanding of the scenarios for working with it. This misunderstanding can be caused by the missing or incomplete documentation accompanying the library, the small number of use cases, and the informal nature of the existing documentation. Integration errors can be quite non-trivial, and the developer needs to spend a significant amount of time to find them.

This article describes the approach developed by the authors that automates the detection of integration errors in a program with external libraries. The approach is based on the use of formal library specifications to describe the behavior of libraries and library functions. To specify valid library usage scenarios, we use LibSL specification language [1, 2] which represents the library as a system of interacting extended finite state machines (EFSM). Based on the mentioned specifications, the enlarged model of library functions is built, capable of detecting errors in using the library. This model is fed to the input of the static analyzer together with the source code of the analyzed application. The static analyzer (we use the KEX analyzer for it [3]) combines the program model and the synthesized library model for joint analysis. At the same time, the library model is formed in such a way that library usage errors (integration errors) are detected by standard analyzer algorithms. The developed approach was implemented as an extension of KEX static analyzer and successfully tested on several Java projects.

The rest of the paper is organized as follows. The second section contains definitions of software integration errors. Section 3 contains short description of LibSL. Section 4 explains the main idea of proposed approach. The 5th section considers the synthesis of library approximation. Section 6 describes the implementation of our approach as a tool called SPIDER. Section 7 acquaints with the evaluation of the developed tool. The 8th section contains a conducted review of the related work. In the conclusion, we present the obtained results and discuss possible future work.

## 2  Integration Errors

Large-scale software systems consisting of many heterogeneous components, as well as other programs, may contain software errors. Some of these errors have the same nature as the errors of simple programs, while others are caused precisely by the difficulties of integrating components with each other. The reason for the integration errors is that the main program violates the protocol of interaction with the library. The protocol of interaction with the library is the rules for calling individual API functions and valid scenarios for calling various API functions.

In this paper, we will consider the following types of software integration errors: violations of the contracts of library API functions and incorrect order of library API function calls.

In software engineering, function contracts define the preconditions and postconditions of functions in the form of logical assertions. The preconditions specify the assertions that must be true when calling the function. The postconditions specify assertions whose truth is guaranteed by the function if the precondition is fulfilled. Violations of API function contracts in multicomponent programs are usually associated with calling functions with incorrect argument values.

Incorrect order of library API function calls are manifested in the fact that API functions are called on some object not in the sequence, in which it is permissible by valid library usage scenarios. For example, a data transfer operation through a stream is called before the stream attributes are set.

## 3   Library Specification Language

To specify the structure and behavior of libraries in this work, we used the previously developed LibSL language [1][2][5]. LibSL is designed to describe the structure of the library, the signature and the enlarged behavior of API functions, as well as to set valid scenarios for using the library. The library and its components are represented by the system of interacting extended finite state machines. The main elements of the specification are presented in Listing 1.

```
libsl "1.0.0";
library Computer version "1.0.0";
types {
  Computer (spider.computer.Computer);
  OS (spider.computer.OS);
  OSName (string);
  Int(int32);
}

automaton spider.computer.Computer : Computer {
  initstate Downed;
  state Boot;
  state OSSelected;
  state OSLoaded;
  finishstate Closed;

  var isMemoryInit: bool = false;
  var isOsLoaded: bool = false;

  shift Downed -> Booted(boot);
  shift Boot -> OSSelected(selectOS);
  shift OSSelected -> OSLoaded(loadOS);
  shift any -> Closed(shutdown);

  fun boot()
    requires isMemoryNotInit: !isMemoryInit;
  {
    isMemoryInit = true;
  }
  fun selectOS(osName: OSName);
  fun setBootPartition(part: Int)
    requires partitionLimits: part >= 0 & part < 16;
  fun loadOS()
    requires isMemoryInit: isMemoryInit;
    requires isOsNotLoaded: !isOsLoaded;
  {
    isOsLoaded = true;
  }
```

```
  fun shutdown() {
    isOsLoaded = false;
    isMemoryInit = false;
  }
}
```

**Listing 1.** Example of LibSL specification

The example contains a fragment of the test library specification and consists of the following descriptions:

- library automaton (*automaton* keyword);
- state of the automaton (*state*, *initstate* and *finishstate* keywords);
- automaton transitions (*shift* keyword);
- automaton internal variables (*var* keyword);
- signatures of library API functions (*fun* keyword) and their aggregated behavior.
- precondition for functions (*requires* keyword);
- annotated semantic types (*types* keyword).

It is assumed that to find possible integration errors, the developer will create a specification of the libraries used exploiting the capabilities of the LibSL language.

## 4 Main Idea

The main idea of integration error detection proposed in this article is to use formal library specifications to synthesize approximations of the behavior of API library functions capable of detecting violations of the library usage protocol. The mentioned approximations are embedded in the code of the testing program in parallel with the source libraries. The resulting code is the input of a static analyzer. The task of detecting integration errors is reduced to solving the problem of the reachability of the approximation code that diagnoses the error.

To achieve the goal of the research, the following tasks are solved:

- definition of integration errors classes detected in the work;
- development of the approximation model code, which can detect all classes of selected integration errors;
- development of the model code generator to integrate error detector into the internal model of the KEX static analyzer;
- development of a new work mode of the KEX static analyzer, which allows analyzing libraries.

The peculiarity of the proposed approach is to combine the library code and the synthesized model approximation code inside the program model. Meanwhile, the model approximation code does not affect the behavior of the source program in any way. Such a solution allows, on the one hand, preserving the original behavior of the testing program, and, on the other hand, detecting a violation of the specification.

The high-level approach scheme is shown in Figure 1. The input of the analyzer is supplied with the source program, the library, and an approximation of the "ideal" behavior of the library synthesized based on the specification. The static analyzer processes input artifacts together and generates an error report. Information from the library and its specification is used for the most detailed diagnosis of errors found. Diagnostic message contains information about the type of the error and its location.

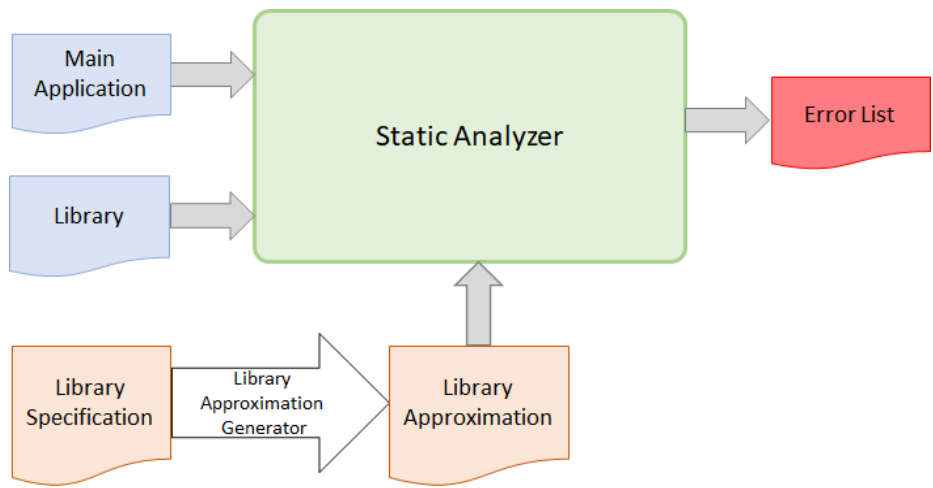

**Fig. 1.** High-level approach scheme

## 5     Synthesis of library approximation

The key idea of the proposed approach is to create an "ideal" approximation of the library behavior based on the specification, which can be used as an oracle to check the correctness of the joint work of the application and the library. In the first version of our approach, the approximation was built as a program code in the Java language, in which all the functions of the source library were generated with the same signatures, but with artificial bodies, in which transitions of automata from one state to another were simulated and checks of contracts and correctness of states were embedded. In the current version, we have abandoned code generation in Java and create approximations immediately in the form of the internal representation of the static analyzer by instrumenting the internal representation of the original library function. This makes the synthesis process more optimal. However, in this section, for ease of understanding the synthesis process, we will operate in terms of elements of the Java language.

Creating an approximation consists of several components. Firstly, based on the automata available in the specification, data structures are created that store the necessary attributes of automata, such as states and internal variables and so on. Secondly, for each API function of the library, a similar function is built in approximation, having the same signature, and the body of which solves the following tasks:

- checking the correctness of the current state of the automaton and signaling an integration errors;
- changing the current state of the automaton;

- checking contract preconditions at the beginning of the function and signaling an integration error in case of violation;
- checking contracts postconditions after the completion of the function and signaling an integration error in case of a violation.

The detection of integration errors is implemented by embedding specialized internal calls of the static analyzer (intrinsic) into the approximation code, the reachability of which in the program indicates an existing integration error. Additional parameters of these calls allow localizing the error location and getting its additional attributes.

## 6    Implementation

We implemented the developed approach as a pilot tool, which we called SPIDER (SPECIFICATION-based Integration Defect Revealer). The tool is designed to analyze Java programs using external libraries. Figure 2 shows the architecture of the developed tool, detailing the approach scheme, which is shown in Figure 1.

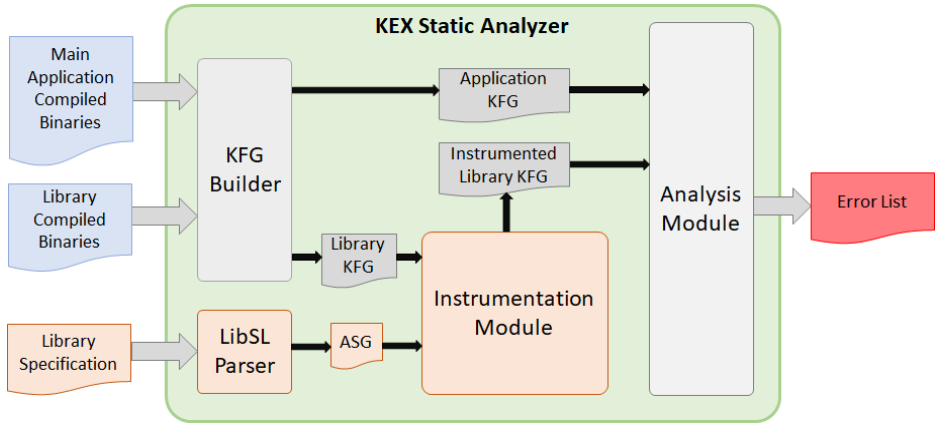

**Fig. 2.** SPIDER Architecture

The input to the tool is the source program, library and specification. Since there may be no source code for external libraries, the tool accepts the compiled bytecode of the library as the input. For unification, the main application is also submitted to the input in the form of compiled bytecode.

As an analyzer we use the static analyzer KEX [3]. This is a powerful and extendable analyzer that uses SMT-solvers to process a code analysis of JVM bytecode (as a result, it can analyze code in many JVM languages, like Java, Kotlin, Scala, etc). KEX uses a special internal representation of JVM programs called KFG, which is essentially a variant of the SSA model implementation. For building KFG KEX uses the library named kfg[1] to represent and to instrument JVM bytecode. Both KFG for the main application and for the library are built using the KFG Builder Module, then the KFG graph is used by the Analysis Module. Library specification is converted to an abstract semantic graph (ASG) in LibSL Parser[2]. The ASG represents information about automata, data types, variables, etc. Instrumentation Module modified library KFG aims to insert library approximation behavior into library model. Analysis Module starts the analysis using the built main application KFG and instrumented library KFG. It converts the program into Predicate State representation [4] and uses z3 SMT-solver for defect detection. Instrumentation Module and Analysis Module were developed as KEX extensions.

---

[1] https://github.com/vorpal-research/kfg

Let us consider SPIDER in detail. The kfg library allows the manipulation of JVM bytecode on a higher level. It allows instrumenting automata in original code classes. We put a state transition code in specified methods. A simple example of such transition obtained by processing of specification from Listing 1 is shown in Listing 2.

```
public void selectOS(String name) {
  if (this.$KEX$INSTRUMENTED$STATE == 2) {
    // State transition code
    this.$KEX$INSTRUMENTED$STATE = 3;
  }
  else {
    boolean[] var16 = new boolean[] { false };
    // intrinsic denotes a forbidden state
    AssertIntrinsics.kexAssert("id2", var16);
  }
  ... // location of original method's code
}

public void setBootPartition(int partition) {
  // precondition calculation
  boolean temp1 = partition >= 0 && partition < 16;
  boolean[] temp2 = new boolean[] { temp1 };
  // intrinsic denotes a precondition violation
  AssertIntrinsics.kexAssert("partitionLimits", temp2);
}
```

**Listing 2**. Fragment of instrumented code

The example shows two main concepts of SPIDER: automaton state with its transition code and contracts. Contracts are implemented as Boolean variables, the value of which is calculated according to the predicate. The result is passed as a parameter to the internal KEX intrinsic function. Automaton state transitions are implemented as three statements: current state check, new state assignment and intrinsic function call. The Analysis Module finds the library API function calls and checks their conditions. If the condition can be false then intrinsic call is reachable, and the error occurs.

---

[2] https://github.com/vorpal-research/libsl

The current version of SPIDER has a limitation. Contracts of functions cannot describe the requirements for complex data types, such as arrays and collections (for example, their size and content). We are planning to fix it later.

## 7  Evaluation

To evaluate our approach, we conducted a series of experiments. For the first experiment, we use a simple artificial library shown in Listing 1. We create a small application that contains all types of integration errors, which can be detected by our approach. The small fragment of this application is shown in Listing 3.

```
// correct library usage
Computer computer1 = new Computer();
computer1.boot();
computer1.setBootPartition(0);
computer1.selectOS("win");
computer1.loadOS();

// wrong sequence of function call
Computer computer2 = new Computer();
computer2.boot();
computer2.loadOS();
computer2.selectOS("linux");
// Error: selectOS called after loadOS

// shift from finish state
Computer computer3 = new Computer();
computer3.shutdown();   // finish state
computer3.boot();
// Error: try to shift from finish state

// precondition violation: partition must be positive
Computer computer4 = new Computer();
computer4.boot();
computer1.setBootPartition(-1);
// Error: contract violation
```

**Listing 3**. Simple test project

The test library (named Computer) has the eponymous class and all methods shown in the listing, but their implementations are beyond the scope of this paper. We emphasize that in our analysis we could not have the source code of the library, we only need its bytecode. The listing shows the four test cases: a correct library protocol usage, a wrong method call ordering, an erroneous shift from state marked as 'finish' in the specification and a precondition violation. All these cases are covered by the analyzer and errors were successfully detected.

For the second group of experiments we used a well-known library named okHttp[3]. It proposes an elegant way to send HTTP requests. For the evaluation of SPIDER, we used one of the demo projects named OkCurl. The project contains one code file that represents a console application that looks like curl.

We created a specification for okHttp[4] and evaluated SPIDER on it. The results of the evaluation are shown below.

The Listing 4 illustrates the code fragment of OkCurl. We can see a command line argument *connect-timeout* and a variable *connectTimeout* that represents it. The program doesn't validate the argument, so this is the place where an error may occur (e.g. if the value is negative). Our specification covers this case: the error successfully found.

```
@Option(names = ["--connect-timeout"],
 description = ["Maximum time allowed for connection"])
var connectTimeout = DEFAULT_TIMEOUT
// ...
if (connectTimeout != DEFAULT_TIMEOUT) {
  // SPIDER detects an error on the next line
  builder.connectTimeout(connectTimeout.toLong(), SECONDS)
}
```

**Listing 4**. Code fragment of OkCurl

In addition to these errors, we have added our own ones. For example, we commented out the line of code that sets the URL of request. Fragment of the code is shown on the Listing 5. SPIDER successfully found this error.

```
val url = url ?: throw IllegalArgumentException()
// request.url(url) this line was commented out
// SPIDER detects an error on the next line
return request.build()
```

**Listing 5**. Code fragment with artificial error

Also, all other errors that were added in the evaluation were successfully found by SPIDER. This means, our approach and tool can be used to detect integration errors in applications using real libraries.

## 8    Related Work

There are many papers devoted to the use of formal specifications for finding errors in complex programs. We will focus on the articles closest to our approach. In our review, we will be interested in the following aspects:

- Which programming languages are the oriented approach?

---

[3] https://square.github.io/okhttp/

[4] https://github.com/vorpal-research/kex/blob/spider/kex- runner/src/test/resources/org/jetbrains /research/kex/spider/okhttp3.lsl

- Is it possible to specify libraries without having their source code?
- Does the approach allow to create specifications without changing the library itself?
- Does the approach allow describing the behavior of the entire library, and not just individual API functions?
- How accurate and sound is the approach?

Some of the analyzed approaches are focused on the C programming language. Hallam, Chelf, Xii and Engler in [6] propose an approach to finding errors in C programs using checkers. Checkers describe the rules for code fragments in a special Metal language. Metal uses an automaton paradigm to describe correctness checks. It allows you to find violations of known correctness rules and automatically derive such rules from the source code. The limitations of the approach are the strict binding to the C language and the inability to describe the semantics of an entire library. A similar approach is used by Microsoft Research in the SLAM project. [7][8][9][10]. The approach uses the Specification Language for Interface Checking (SLIC) to specify the external API and perform static code analysis to verify the correctness of use of the API. SLIC expresses temporal API security requirements, such as the order of function calls, and the goal of the project is to automatically check whether these requirements are met by the client and the API developer. Complex checks can be described imperatively as insertions in the C language. The resulting code written in SLIC is also compiled into C code. The main goal of the SLAM project is to check the correctness of the drivers. The main limitations of the project are the C programming language used and unsafe and uncontrolled imperative inserts. The WYSIWEB project [11] describes the implementation of a tool that uses a C-like language to search the Linux kernel source code for a certain type of errors and API uses.

Some approaches are initially focused on the Java programming language. The main language for the specification of Java programs is the JML (Java Modeling Language) [12]. It allows you to specify class methods in detail in terms of pre- and postconditions, embedding specifications directly into the code in the form of annotating comments. There are projects for implementing static JML-based analyzers, for example, ESC/Java [13], which statically check JML specifications. The limitations of the approach are the inability to annotate external components without affecting their source code, as well as the lack of an explicit mechanism for describing the correct behavior of the entire external library. Reiss in [14] suggests an approach for finding errors in multicomponent Java programs. The authors have developed a CHET system that allows you to check the correctness of the use of components by checking a variety of conditions. CHET is implemented as an interactive tool that allows the programmer to visually monitor the correctness of working with components.

Bierhoff in his dissertation [15] reveals the issues of compliance with API protocols in object-oriented software. Two key entities made it possible to implement the proposed approach: object-oriented types of states defined by means of hierarchical state refinements and access permissions. Ferles, Stephens and Dillig in [16] propose to set API specifications using parameterized context-summary grammars. This approach does not have enough power to set the behavior of complex libraries. A group of researchers from University de Montreal in [17] presented the results of a study of several types of API usage restrictions, as well as their documentation. The authors have developed an algorithm that automatically determines four types of usage restrictions in the API source code.

In the approaches discussed earlier, the specifications were created manually, but there is a group of approaches in which the specifications are extracted from existing programs. Thus, in [18], a method of mining specifications from external components is proposed, followed by static checking of the correctness of the implementation of this specification. The method is very promising, but its significant limitations: the need for the source code of the library, as well as the low accuracy and soundness of the obtained specifications and, as a result, the possible omission of errors or false positives detections. A similar approach to dynamically extracting library usage protocols is offered by Pradel from ETH [19]. Dynamic analysis is used to analyze the instrumented program and extract the automatic specification. It is further refined and can be used to test other programs. The limitations of the method are like the previous one: low completeness and accuracy of the approach.

## 9 Conclusion

The paper describes the approach developed by the authors to detect integration errors based on the use of formal library specifications in the LibSL language. Formal specifications are used to build approximations of libraries that are used as oracles for the program being analyzed. Based on the approach, the SPIDER tool was developed, implemented as an extension of the KEX static analyzer. Experiments on simple projects with artificial and industrial libraries have shown the fundamental applicability of the approach and the operability of the tool.

The directions of further research are related to the expansion of the apparatus for describing library functions, including support for complex data types; expansion of classes of detected integration errors; support for different modes of joint analysis of libraries and their approximations.

The areas of improvement of the tool are related to overcoming the existing technical limitations of the tool, creating Java Standard Library specifications, conducting more experiments with large-scale industrial projects using popular open libraries.

## References

1. Itsykson V. M. LibSL: Language for Specification of Software Libraries, Programmnaya Ingeneria, 2018, vol. 9, no. 5, pp. 209—220.
2. Itsykson, V.M. Formalism and Language Tools for Specification of the Semantics of Software Libraries. *Aut. Control Comp. Sci.* **51,** 531–538 (2017). https://doi.org/10.3103/S0146411617070100
3. A. Abdullin, M. Akhin and M. Belyaev, "Kex at the 2021 SBST Tool Competition," *2021 IEEE/ACM 14th International Workshop on Search-Based Software Testing (SBST)*, 2021, pp. 32-33, doi: 10.1109/SBST52555.2021.00014
4. Akhin M., Belyaev M., Itsykson V. (2017) Borealis Bounded Model Checker: The Coming of Age Story. In: Mazzara M., Meyer B. (eds) Present and Ulterior Software Engineering. Springer, Cham. https://doi.org/10.1007/978-3-319-67425-4_8
5. V. Itsykson. Partial Specifications of Libraries: Applications in Software Engineering / TMPA-2019, November 2019, CCIS, volume 1288, pp 3-25

6. Seth Hallem, Benjamin Chelf, Yichen Xie, and Dawson Engler. 2002. A system and language for building system-specific, static analyses. *SIGPLAN Not.* 37, 5 (May 2002), 69–82. DOI:https://doi.org/10.1145/543552.512539
7. Thomas Ball and Sriram K. Rajamani. 2002. The SLAM project: debugging system software via static analysis. *SIGPLAN Not.* 37, 1 (Jan. 2002), 1–3. DOI:https://doi.org/10.1145/565816.503274
8. Thomas Ball and Sriram K. Rajamani. 2001. Automatically validating temporal safety properties of interfaces. In *Proceedings of the 8th international SPIN workshop on Model checking of software* (*SPIN '01*). Springer-Verlag, Berlin, Heidelberg, 103–122.
9. Ball, T., Bounimova, E., Levin, V., Kumar, R., & Lichtenberg, J. (2010). The Static Driver Verifier Research Platform. *CAV*.
10. Thomas Ball, Vladimir Levin, and Sriram K. Rajamani. 2011. A decade of software model checking with SLAM. *Commun. ACM* 54, 7 (July 2011), 68–76. DOI:https://doi.org/10.1145/1965724.1965743
11. J. L. Lawall, J. Brunel, N. Palix, R. R. Hansen, H. Stuart and G. Muller, "WYSIWIB: A declarative approach to finding API protocols and bugs in Linux code," *2009 IEEE/IFIP International Conference on Dependable Systems & Networks*, 2009, pp. 43-52, doi: 10.1109/DSN.2009.5270354.
12. Gary T. Leavens, Albert L. Baker, and Clyde Ruby. 2006. Preliminary design of JML: a behavioral interface specification language for java. *SIGSOFT Softw. Eng. Notes* 31, 3 (May 2006), 1–38. DOI:https://doi.org/10.1145/1127878.1127884
13. Cormac Flanagan, K. Rustan M. Leino, Mark Lillibridge, Greg Nelson, James B. Saxe, and Raymie Stata. 2002. Extended static checking for Java. *SIGPLAN Not.* 37, 5 (May 2002), 234–245. DOI:https://doi.org/10.1145/543552.512558
14. Steven P. Reiss. 2005. Specifying and checking component usage. In *Proceedings of the sixth international symposium on Automated analysis-driven debugging* (*AADEBUG'05*). Association for Computing Machinery, New York, NY, USA, 13–22. DOI:https://doi.org/10.1145/1085130.1085133
15. Bierhoff, K.: API Protocol Compliance in Object-Oriented Software. PhD thesis, Carnegie Mellon University, School of Computer Science (April 2009)
16. Kostas Ferles, Jon Stephens, and Isil Dillig. 2021. Verifying correct usage of context-free API protocols. *Proc. ACM Program. Lang.* 5, POPL, Article 17 (January 2021), 30 pages. DOI:https://doi.org/10.1145/3434298
17. M. Pradel, C. Jaspan, J. Aldrich and T. R. Gross, "Statically checking API protocol conformance with mined multi-object specifications," *2012 34th International Conference on Software Engineering (ICSE)*, 2012, pp. 925-935, doi: 10.1109/ICSE.2012.6227127.
18. Michael Pradel. 2009. Dynamically inferring, refining, and checking API usage protocols. In *Proceedings of the 24th ACM SIGPLAN conference companion on Object oriented programming systems languages and applications* (*OOPSLA '09*). Association for Computing Machinery, New York, NY, USA, 773–774. DOI:https://doi.org/10.1145/1639950.1640008
19. M. A. Saied, H. Sahraoui and B. Dufour, "An observational study on API usage constraints and their documentation," *2015 IEEE 22nd International Conference on Software Analysis, Evolution, and Reengineering (SANER)*, 2015, pp. 33-42, doi: 10.1109/SANER.2015.7081813.